**Title:** Generalized Design of Basket Trials with P-value Combination Test

Heng Zhou*, Linda Sun, Fang Liu, Cong Chen

Biostatistics and Research Decision Sciences, Merck & Co., Inc., Rahway, NJ 07065, USA

*Corresponding author: Heng Zhou, MAILSTOP UG-1CD44, 351 North Sumneytown Pike, North Wales, PA 19454, USA, heng.zhou@merck.com

**Abstract**:

The oncology exploratory basket trial design with pruning and pooling (P&P) approach has gained increasing popularity in recent years for its simplicity and efficiency. This method was proposed based on binary endpoint, limiting its wider application. This short communication proposed a generalized framework of using P-value combination test to implement pruning and pooling process in basket trials. Only P-values of any type of statistical testing from each cohort are needed for decision making, which provides great flexibility for basket trial designs with P&P approach.

The pruning and pooling (P&P) approach is relatively a recent development that aims to improve the efficiency of oncology exploratory basket clinical trials [1-4]. It is a straightforward approach to borrow information across different indications (or any pre-defined subgroups) in the basket trials and can efficiently test if the drug is working in any of the indications, which is the major goal in the exploratory phase of oncology drug development. The P&P approach is usually implemented in a two-step manner. In the pruning step, inactive indications would be pruned given specific pruning bars with independent analysis on each individual indication. Then in the pooling step, the data across the indications which were not pruned in the first step will be pooled to conduct the final analysis at a significance level $\alpha^*$ in order to control the overall type I error rate at $\alpha$, where $\alpha^*$ is usually smaller than $\alpha$ because the pooled test should pay some penalty for "cherry-picking" in the first step.

Current methods of basket trial designs with P&P approach mainly focus on the binary endpoint, where the pruning bars in the first step are calculated as the required number of responses in each indication and the pooled test level $\alpha^*$ is also derived from the minimal number of responses across all the pooled indications to claim significance. Nevertheless, adapting this approach to other endpoints, such as continuous or time-to-event endpoints, presents significant challenges. Meanwhile, as the historical data plays more important roles in the exploratory single-arm studies, we can test the data in each cohort with some covariate-adjustment approach, where a set of P-values are available for decision-making. To provide a general framework to accommodate all types of endpoints and trial settings, we can consider using P-value combination test to implement the idea of P&P in the exploratory basket trials. Specifically, a threshold of P-values is set at the pruning step and only the indications with individual test P-values less than the threshold will be pooled for the final analysis by combining those P-values. In this way, the basket trial designs with P&P approach can be generalized as a truncated P-value combination test process, which could be implemented in any exploratory basket trials with P-values from each indication are available.

There are various methods of P-value combination test, among which the inverse normal combination test [5] and Fisher's combination test [6] are most widely used. Here we use the inverse normal combination test to demonstrate how it is used in basket trials, as it is arguably more powerful than the Fisher's test [7]. We actually explored both and observed similar operating characteristics only when sample sizes are comparable across the cohorts.

Let's consider a basket trial with $K$ test indications and the P-values of independent analysis for each indication are $p_k, k = 1, \cdots, K$. Suppose the threshold of P-values in the pruning step is $\tau$, then the test statistic for final pooled analysis is:

$$W_K(\tau) = \sum_{k=1}^{K} \omega_k \, \Phi^{-1}(1 - p_k) \, I(p_k \leq \tau), \tag{1}$$

where $\omega_k$ is the weights of each indication and $\sum_{k=1}^{K} \omega_k^2 I(p_k \leq \tau) = 1$. Notice that here we just consider the one-stage basket trial, which means no more patients will be enrolled in the pooling step, so that the P-values less than the threshold could be directly used in the pooling analysis. Then, to control the type I error rate at $\alpha$ level under the null hypothesis $H_0$ that the drug does not work in any of the indications, we need to derive for the critical value $w^*$ such that $P(W_K(\tau) > w^* | H_0) = \alpha$. This could be achieved by numerical study with the condition that $p_k \overset{i.i.d}{\sim} Uniform(0,1)$ under the null hypothesis. To better understand the critical value $w^*$, we can map it to the standard normal test and get the significance level $\alpha^* = 1 - \Phi(w^*)$ to provide some intuition about the magnitude of the penalty paid at the final analysis, although the test statistic is no longer following standard normal distribution. Smaller $\alpha^*$ indicates more stringent testing with more penalty paid. Figure 1 shows the significance level of final pooled analysis given the P-value threshold in the pruning step under different numbers of indications, controlling the overall type I error rate at $\alpha = 0.05$ level. And we use the equal weights of each indication in test statistic $W_K(\tau)$. It is as expected that $\alpha^*$ is smaller with more indications in the basket trial as more inactive indications would be potentially included in the final analysis. Within the same number of indications, we observe a lowest plateau of $\alpha^*$ around P-value thresholds from 0.15 to 0.25. This pattern is as expected, as for smaller P-value thresholds, the pruning step is sufficiently stringent, so less penalty is needed to be paid at final analysis as the type I error mainly comes from the pruning step. And for larger P-value thresholds approaching to 1, the test statistic would be approximately standard normal so the testing level would be approaching to the nominal $\alpha$ level.

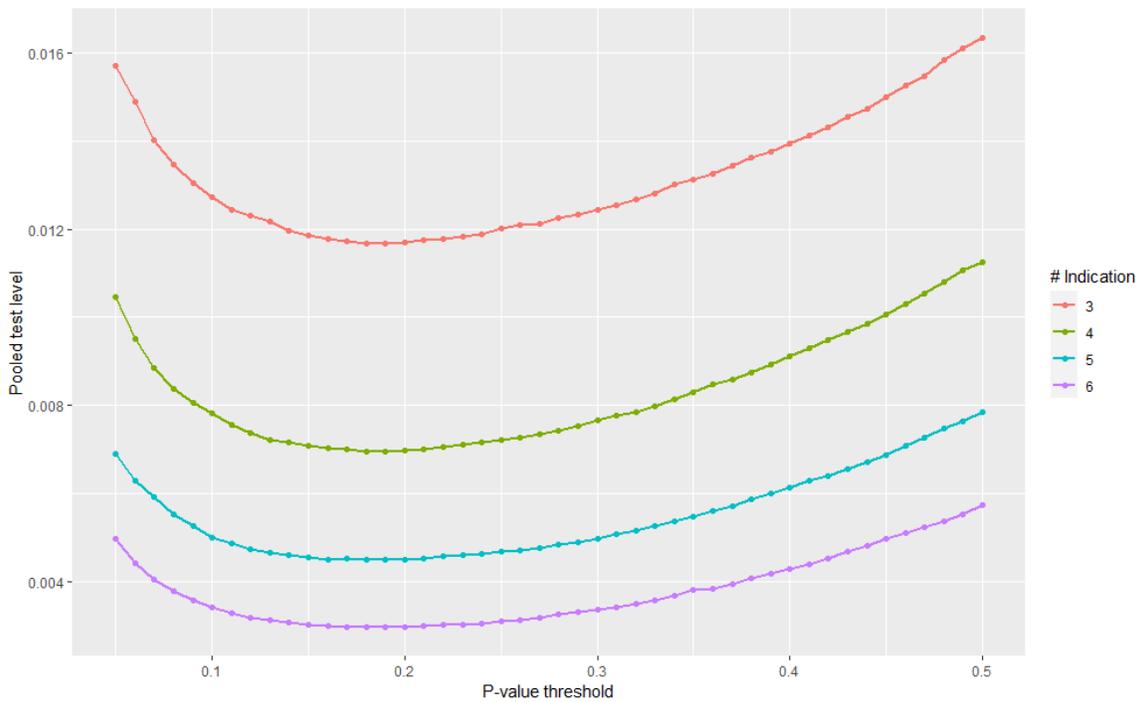

Figure 1. Critical value vs. P-value threshold to control the type I error rate 0.05.

To evaluate the power of P-value combination test in the basket trial, we can firstly calculate the probability of rejecting the null hypothesis $p(G)$ given specific number of truly active indications $G = 1, \cdots, K$, respectively. Suppose we do not have prior knowledge on how many truly active indications, then the overall power can be defined as the average of $p(G)$ over $G = 1, \cdots, K$. In fact, if $G$ follows certain distribution given prior information, the overall power can be calculated accordingly. We use simulation to get the power and generate the P-values of alternative hypothesis as $\Phi[\Phi^{-1}(1 - U) - \gamma]$. Here, $U$ is randomly drawn from $Uniform(0,1)$ and $\gamma$ represents the treatment effect that the test statistics under alternative hypothesis is assumed to follow $N(\gamma, 1)$ for each cohort. In our simulation, we set $\gamma = 2$ which translates the 63% individual power of each cohort. Figure 2 shows the curves of power given P-value threshold under different number of indications in the basket trial. For simple illustration, we consider same value of $\gamma$ in the simulation, while in practice we can generate different treatment effect in the simulation to evaluate the power per trial setting. As shown in the Figure 2, the power reaches to a maximum plateau around P-value thresholds from 0.15 to 0.25, which is the same range yielding lowest plateau observed in Figure 1. This is assuring that we do not need to worry about loss of efficiency when paying comparatively high penalty. This could serve as a recommended range of P-value thresholds in the pruning step in real practice.

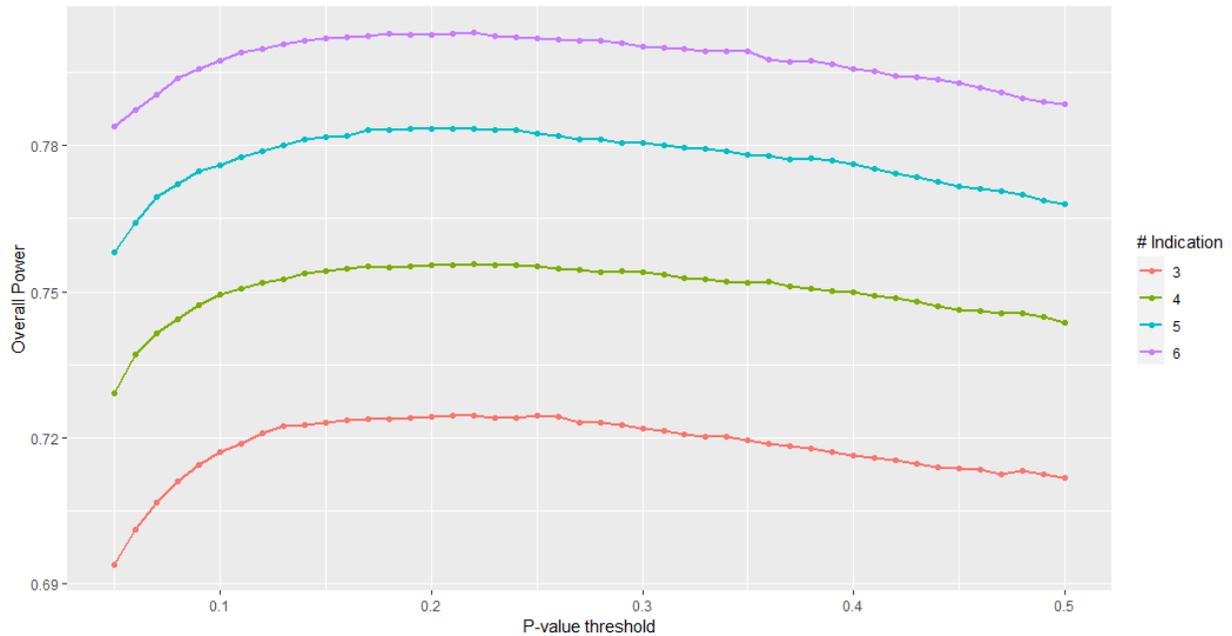

Figure 2. Overall power vs. P-value threshold; type I error rate is controlled at 0.05.

We can implement the Fisher's combination test similarly as above in the basket trials. Actually the truncated Fisher's combination test has been studied extensively [8,9]. In addition to the fact that the inverse normal combination test could achieve more efficiency, there are two other favorable considerations compared to Fisher's combination test: 1) The Fisher's test statistic critical values are not straightforward to interpret, as the Chi-square distribution has different degrees of freedom given different numbers of indications in the pooling analysis; 2) Although we mainly focus on the one-stage basket trial in the manuscript, it is not difficult to extend the inverse combination test to two-stage trial where we will get a new set of P-values with more patients enrolled at the second stage. Similar practice has been implemented in various adaptive designs, with some adjustment of P-values used in the combination test may be needed [10,11]. However, the Fisher's combination test may not embrace such flexibility.

In this manuscript, we use the equal weights across the cohorts in the test statistic of inverse normal combination test. In real practice, we can assign different weights according to the priority of each indication in clinical development. Considering it is challenging to determine the priority quantitatively, a feasible way is to assign the weights as $\omega_k = \sqrt{n_k/sum(n_k)}$, where $n_k$ is the adjusted sample size for each cohort in the pooled analysis.

**References**

1. Zhou H, Liu F, Wu C, Rubin EH, Giranda VL, Chen C. (2019) Optimal two-stage designs for exploratory basket trials. *Contemporary Clinical Trials*, 85, p.105807.

2. Wu C, Liu F, Zhou H, Wu X, Chen C. (2021) Optimal one-stage design and analysis for efficacy expansion in Phase I oncology trials. *Clinical Trials*, 18(6), 673-680.

3. Wu X, Wu C, Liu F, Zhou H, Chen C. (2021) A Generalized Framework of Optimal Two-stage Designs for Exploratory Basket Trials. *Statistics in Biopharmaceutical Research*, 13(3), 286-294.

4. Jing N, Liu F, Zhou H and Chen C. (2022) An optimal two-stage exploratory basket trial design with aggregated futility analysis. *Contemporary Clinical Trials*, 116, p.106741.

5. Stouffer S, DeVinney L, Suchmen E. (1949) The American soldier: Adjustment during army life. Vol.1. *Princeton University Press*; Princeton, US.

6. Fisher RA. (1932) Statistical Methods for Research Workers. Oliver and Boyd, Edinburgh.

7. Whitlock MC. (2005) Combining probability from independent tests: the weighted Z-method is superior to Fisher's approach. *Journal of Evolutionary Biology*, 18(5), 1368-1373.

8. Zaykin DV, Zhivotovsky LA, Westfall PH and Weir BS (2002). Truncated product method for combining P-values. *Genetic Epidemiology*, 22, 170–185.

9. Zhang H, Tong T, Landers J, Wu Z. (2020). TFisher: a powerful truncation and weighting procedure for combining p-values. *The Annals of Applied Statistics*, 14(1), 178-201.

10. Lehmacher W, Wassmer G. (1999) Adaptive sample size calculations in group sequential trials. *Biometrics*, 55:1286–1290.

11. Jorgens S, Wassmer G, Konig F, Posch M. (2017) Nested combination tests with a time-to-event endpoint using a short-term endpoint for design adaptations. *Pharmaceutical Statistics*, 18(3), 329-350.

Appendix: R code to implement truncated inverse normal combination test in basket trials

```
### Get type I error rate
T1E.TInvnorm <- function(K, p.cut, alphastar, nsim=100000, seed=43)
{
 #K: number of indications in the basket trial
 #p.cut: the P-value threshold for pruning
 #alphastar: testing level at final analysis
 #nsim: number of simulations

 set.seed(seed)
 sig=0

 for (sim in 1:nsim)
 {
  p.values <- 1-runif(K)
  p.values <- p.values[p.values<=p.cut]
  if(length(p.values)>0)
  {
   w <- sum(sqrt(1/length(p.values))*qnorm(1-p.values))

   sig = sig + as.numeric(w>qnorm(1-alphastar))
  }

 }
 return(sig/nsim)
}

### Function for solving alphastar to yiled target type I error rate level alpha0
diff0.TInvnorm <- function(alpha0, K, p.cut, nsim, seed, alphastar) {
 T1E.TInvnorm(K, p.cut, alphastar, nsim, seed) - alpha0
}

### Get power with fixed number of truly active indications
Power.TInvnorm <- function(K, p.cut, alpha, G, gamma, nsim=100000, seed=43) {

 #K: number of indications in the basket trial
 #p.cut: the P-value threshold for pruning
 #alpha: overall type I error rate
 #G: number of truly active indications
 #gamma: the parameter to generate P-values under alternative hypothesis
 #nsim: number of simulations

 alphastar <- uniroot(diff0.TInvnorm, alpha0=alpha, K=K, p.cut=p.cut, nsim=nsim, seed=seed, c(0,1))$root
```

```r
sig = 0
set.seed(seed)
for (sim in 1:nsim)
{
  p.null <- 1-runif(K-G)
  p.alt <- pnorm(qnorm(1-runif(G)) - gamma)
  p.values <- c(p.null,p.alt)[c(p.null,p.alt)<=p.cut]

  if(length(p.values)>0)
  {
    w <- sum(sqrt(1/length(p.values))*qnorm(1-p.values))

    sig = sig + as.numeric(w>qnorm(1-alphastar))
  }

}
sig/nsim
}
```